\documentclass[12pt]{article}

\usepackage{amsfonts}			
\newcommand{\RR}{{\mathbb R}}
\newcommand{\CC}{{\mathbb C}}
\newcommand{\HH}{{\mathbb H}}
\newcommand{\OO}{{\mathbb O}}

\newcommand\cyr{}\newcommand\EE{}

\newcommand{\tr}{{\rm tr}}		

\renewcommand{\bar}{\overline}
\renewcommand{\tilde}{\widetilde}
\renewcommand\AA{{\cal A}}
\newcommand\BB{{\cal B}}
\renewcommand\EE{{\cal E}}
\newcommand\II{I}
\newcommand\MM{{\cal M}}
\newcommand\PP{{\cal P}}
\newcommand\QQ{{\cal Q}}
\newcommand\VV{{\cal V}}

\begin{document}



\title{
{\bfseries THE EXCEPTIONAL JORDAN EIGENVALUE PROBLEM}
\footnote{Internat.\ J.\ Theoret.\ Phys.\ (1999; to appear)}
}

\author{Tevian Dray \\
	{\it Department of Mathematics, Oregon State University,
		Corvallis, OR  97331} \\
	{\tt tevian{\rm @}math.orst.edu} \\
		\and
	Corinne A. Manogue \\
	{\it Department of Physics, Oregon State University,
		Corvallis, OR  97331} \\
	{\tt corinne{\rm @}physics.orst.edu}
}

\date{\normalsize (30 March 1999; last revised 31 October 1999)}

\maketitle

\begin{abstract}
We discuss the eigenvalue problem for $3\times3$ octonionic Hermitian matrices
which is relevant to the Jordan formulation of quantum mechanics.  In contrast
to the eigenvalue problems considered in our previous work, all eigenvalues
are real and solve the usual characteristic equation.  We give an elementary
construction of the corresponding eigenmatrices, and we further speculate on
a possible application to particle physics.
\end{abstract}


\section{\textbf{Introduction}}

In previous work~\cite{Eigen,Find,NonReal} we considered both the {\it left\/}
and {\it right\/} eigenvalue problems for $2\times2$ and $3\times3$ octonionic
Hermitian matrices, given explicitly by
\begin{equation}
\label{Left}
\AA \, v = \lambda \, v
\end{equation}
and
\begin{equation}
\label{Right}
\AA \, v = v \, \lambda
\end{equation}
respectively.  We showed in~\cite{Eigen} that the left eigenvalue problem
admits nonreal eigenvalues over both the quaternions $\HH$ and the octonions
$\OO$, while the right eigenvalue problem admits nonreal eigenvalues only over
$\OO$.  Some of the intriguing properties of the eigenvectors corresponding to
these nonreal eigenvalues were considered in~\cite{NonReal}, and
in~\cite{Dim,Spin} we discussed possible applications to physics, including
the remarkable fact that simultaneous eigenvectors of all 3 angular momentum
operators exist in this context.

However, the main result in~\cite{Eigen} concerned real eigenvalues in the
$3\times3$ octonionic case.  For this case, there are 6, rather than 3, real
eigenvalues~\cite{Ogievetsky}.  We showed that these come in 2 independent
families, each consisting of 3 real eigenvalues which satisfy a modified
characteristic equation rather than the usual one.  Furthermore, the
corresponding eigenvectors are not orthogonal in the usual sense, but do
satisfy a generalized notion of orthogonality (see
also~\cite{Find,Rochester}).  Finally, all such matrices admit a decomposition
in terms of (the ``squares'' of) orthonormal eigenvectors.  However, due to
associativity problems, these matrices are {\it not\/} idempotents (matrices
which square to themselves).

It is the purpose of this paper to describe a related eigenvalue problem for
$3\times3$ Hermitian octonionic matrices which does have the standard
properties: There are 3 real eigenvalues, which solve the usual characteristic
equation, and which lead to a decomposition in terms of orthogonal
``eigenvectors'' which are indeed (primitive) idempotents.

This is accomplished by considering the {\it eigenmatrix\/} problem
\begin{equation}
\label{JEigen}
\AA \circ \VV = \lambda \VV
\end{equation}
where $\VV$ is itself an octonionic Hermitian matrix and $\circ$ denotes the
{\it Jordan product}~\cite{Jordan,JNW}
\begin{equation}
\label{JProd}
\AA \circ \BB = {1\over2} \left( \AA\BB + \BB\AA \right)
\end{equation}
which is commutative but not associative.  We further restrict $\VV$ to be a
(primitive) idempotent; as discussed below, this ensures that the Jordan
eigenvalue problem~(\ref{JEigen}) reduces to the traditional eigenvalue
problem~(\ref{Right}) in the non-octonionic cases.

The exceptional Jordan algebra of $3\times3$ octonionic Hermitian matrices
under the Jordan product, now known as the Albert algebra, was extensively
studied by Freudenthal~\cite{Oktaven,Freudenthal,FreudEng},~%
\footnote{Freudenthal's early work on this topic was originally distributed
in German in mimeographed form~\cite{Oktaven}, parts of which were later
summarized in~\cite{Freudenthal}, which we henceforth cite.  Many of these
results can also be found in English in~\cite{FreudEng}.}
and is well-known to mathematicians~\cite{Jacobson,Schafer,Harvey,Rosenfeld}.
In particular, the existence of a decomposition in terms of orthogonal
idempotents, and its relationship to the eigenvalue problem~(\ref{JProd}), was
shown already in~\cite{JNW}.  Furthermore, since any Jordan matrix can be
diagonalized by an $F_4$ transformation~\cite{Freudenthal}, and since $F_4$ is
the automorphism group of the Jordan product~\cite{Chevalley}, the eigenmatrix
problem~(\ref{JEigen}) is easily solved in theory.  However, we are not aware
of an elementary treatment along the lines presented here.

Our motivation for studying this problem is the well-known fact that the
Albert algebra is the only exceptional realization of the Jordan formulation
of quantum mechanics~\cite{Jordan,JNW,Albert,GT}; over an associative division
algebra, the Jordan formalism reduces to standard quantum mechanics.
Furthermore, the 4 division algebras $\RR$, $\CC$, $\HH$, and $\OO$ are
fundamentally associated with the Killing/Cartan classification of Lie
algebras --- corresponding to physical symmetry groups --- into orthogonal,
unitary, symplectic, and exceptional types.  This most exceptional quantum
mechanical system over the most exceptional division algebra provides an
intriguing framework to study the basic symmetries of nature.

We begin by summarizing the properties of the Albert algebra in
Section~\ref{ALBERT}.  In order to make our work accessible to a wider
audience, we first motivate our subsequent computation by briefly reviewing
the Jordan formulation of quantum mechanics in Section~\ref{JQM}, before
presenting the mathematical details of the eigenvalue results in
Section~\ref{JORDAN}.  In Section~\ref{DISCUSSION}, we include a brief but
suggestive discussion of possible applications, such as its relevance for our
recent work on dimensional reduction~\cite{Dim,Spin}.  Finally, in the
Appendix, we show explicitly how to diagonalize a generic Jordan matrix using
$F_4$ transformations.

\section{\textbf{The Albert Algebra}}
\label{ALBERT}

We consider the {\it Albert algebra\/} consisting of $3\times3$ octonionic
Hermitian matrices, which we will call {\it Jordan matrices}.
\footnote{For a review of the basic properties of the octonions, see for
instance~\cite{Eigen} or~\cite{GT,Okubo}.}
The Jordan product~(\ref{JProd}) of two such matrices is commutative but not
associative.  We have in particular that
\begin{equation}
\AA^2 \equiv \AA \circ \AA
\end{equation}
and we {\it define\/}
\begin{equation}
\AA^3 := \AA^2 \circ \AA \equiv \AA \circ \AA^2
\end{equation}
which differs from the cube of $\AA$ using ordinary matrix multiplication.
Other operations on Jordan matrices are the {\it trace}, denoted as usual by
$\tr(\AA)$, and the {\it Freudenthal product}~\cite{Freudenthal}
\begin{equation}
\AA*\BB = \AA \circ \BB - {1\over2} \Big(\AA\,\tr(\BB)+\BB\,\tr(\AA)\Big)
	  + {1\over2} \Big(\tr(\AA)\,\tr(\BB)-\tr(\AA\circ \BB)\Big) \,\II
\end{equation}
where $\II$ denotes the identity matrix and with the important special case
\begin{equation}
\label{FreudSq}
\AA*\AA = \AA^2 - (\tr\,\AA) \, \AA + \sigma(\AA) \, \II
\end{equation}
where
\begin{equation}
\label{Sigma}
\sigma(\AA) =
{1\over2} \left( (\tr\,\AA)^2 - \tr (\AA^2) \right)
\equiv \tr(\AA*\AA)
\end{equation}
There is also {\it trace reversal}
\begin{equation}
\label{TID}
\tilde \AA = \AA -\tr(\AA)\,\II \equiv -2 \, \II*\AA
\end{equation}
and, finally, the {\it determinant}
\begin{equation}
\label{Det}
\det(\AA) = {1\over3} \, \tr \Big( (\AA*\AA) \circ \AA \Big)
\end{equation}
which can equivalently be defined by
\begin{equation}
\label{FDet}
\Big( (\AA*\AA) \circ \AA \Big) = (\det \AA) \, \II
\end{equation}
Expanding~(\ref{FDet}) using~(\ref{FreudSq}), we obtain the remarkable result
that Jordan matrices satisfy the usual characteristic
equation~\cite{Freudenthal}
\begin{equation}
\label{AChar}
\AA^3 - (\tr\,\AA) \, \AA^2 + \sigma(\AA) \, \AA - (\det \AA) \, \II = 0
\end{equation}

Explicitly, a Jordan matrix can be written as
\begin{equation}
\label{Three}
\AA = \pmatrix{p& a& \bar{b}\cr \bar{a}& m& c\cr b& \bar{c}& n\cr}
\end{equation}
with $p,m,n\in\RR$ and $a,b,c\in\OO$, where the bar denotes octonionic
conjugation.  The definitions above then take the concrete form
\begin{eqnarray}
\label{ThreeEq}
\tr\,\AA &=& p + m + n \nonumber\\
\sigma(\AA) &=& pm + mn + pn - |a|^2 - |b|^2 - |c|^2 \\
\det \AA &=& pmn + b(ac) + \bar{b(ac)} - n|a|^2 - m|b|^2 - p|c|^2 \nonumber
\end{eqnarray}

The {\it Cayley plane}, also called the {\it Moufang plane}, consists of those
Jordan matrices $\VV$ which satisfy the restriction~\cite{FreudEng,Harvey}
\begin{equation}
\label{Cayley}
\VV \circ \VV = \VV; \qquad \tr\,\VV = 1
\end{equation}
We will see below that elements of the Cayley plane correspond to projection
operators in the Jordan formulation of quantum mechanics.  As shown
in~\cite{Harvey}, the conditions~(\ref{Cayley}) force the components of $\VV$
to lie in a {\it quaternionic\/} subalgebra of $\OO$ (which depends on $\VV$).
Basic (associative) linear algebra then shows that each element of the Cayley
plane is a primitive idempotent (an idempotent which is {\it not\/} the sum of
other idempotents), and can be written as
\begin{equation}
\label{ISq}
\VV = vv^\dagger
\end{equation}
where $v$ is a 3-component octonionic column vector, whose components lie in
the quaternionic subalgebra determined by $\VV$, and which is normalized by
\begin{equation}
\label{Norm}
v^\dagger v = \tr\,\VV = 1
\end{equation}
Note that $v$ is unique up to a {\it quaternionic\/} phase.  Furthermore,
using (\ref{FreudSq}) and its trace~(\ref{Sigma}), it is straightforward to
show that, for any Jordan matrix $\BB$,
\begin{equation}
\label{ISqF}
\BB*\BB = 0 \Longleftrightarrow \BB \circ \BB = (\tr\,\BB) \, \BB
\end{equation}
which agrees with~(\ref{Cayley}) up to normalization, and which is therefore
the condition that that $\pm \BB$ can be written in the form~(\ref{ISq})
(without the restriction~(\ref{Norm})).  Note further that for any Jordan
matrix satisfying (\ref{ISqF}), the normalization $\tr\,\BB$ can only be zero
if $v$, and hence $\BB$ itself, is zero, so that
\begin{equation}
\label{NormZero}
\BB*\BB = 0 = \tr\,\BB \Longleftrightarrow \BB=0
\end{equation}
since the converse is obvious.

We will need the following useful identities
\begin{eqnarray}
  (\AA*\AA)*(\AA*\AA) &=& (\det \AA) \, \AA \label{Springer}\\
  (\tilde \AA\circ \AA)\circ(\AA*\AA) &=& (\det \AA) \, \tilde \AA \label{ID}
\end{eqnarray}
for any Jordan matrix $\AA$, which can be verified by direct computation.
Finally, we also have the remarkable fact that
\begin{equation}
\label{OCross}
\AA*\AA=0=\BB*\BB \Longrightarrow (\AA*\BB)*(\AA*\BB)=0
\end{equation}
which follows by polarizing~(\ref{Springer})
\footnote{The necessary fact that $\det(\AA+\BB)=0$ follows from the definition
(\ref{Det}) of the determinant in terms of the triple product, the cyclic
properties of the trace of the triple product, and the assumptions on $\AA$
and $\BB$.}
and which ensures that the set of Jordan matrices satisfying (\ref{ISqF}),
consisting of all real multiples of elements of the Cayley plane, is closed
under the Freudenthal product.

Before proceeding further it is illuminating to consider the restriction to
{\it real\/} column vectors.  If $u,v,w\in\RR^3$, then
\begin{eqnarray}
2 \, uu^\dagger \circ vv^\dagger &=& (u\cdot v) \, (uv^\dagger + vu^\dagger) \\
\tr(uu^\dagger \circ vv^\dagger) &=& (u\cdot v)^2
\end{eqnarray}
where $\cdot$ denotes the usual dot product (and where the Hermitian conjugate
of a real matrix is of course just its transpose).  We also have
\begin{equation}
2 \, uu^\dagger * vv^\dagger = (u\times v)(u\times v)^\dagger
\end{equation}
where $\times$ denotes the usual cross product.  We can therefore view the
Jordan product as a generalization of the (square of the) dot product, and the
Freudenthal product as a generalization of the (square of the) cross product.

This somewhat simplified perspective is nevertheless extremely useful in
grasping the essential content of the corresponding octonionic manipulations.
For instance, the linear independence of (real) $u$, $v$, $w$ is given by the
condition
\begin{equation}
\det(Q) =  u\cdot(v\times w) \ne 0
\end{equation}
where $Q$ is the matrix whose columns are the vectors $u$, $v$, $w$.  Note
that
\begin{equation}
QQ^\dagger \equiv uu^\dagger+vv^\dagger+ww^\dagger
\end{equation}
and of course $\det(QQ^\dagger)=|\det(Q)|^2$.  But using the
definition~(\ref{Det}) for real $u$, $v$, $w$ leads to the identity
\begin{equation}
\det(uu^\dagger+vv^\dagger+ww^\dagger) = \Bigl(u\cdot(v\times w)\Bigr)^2
\end{equation}
which not only emphasizes the role played by the determinant in determining
linear independence, but also makes plausible the cyclic nature of the trace
of the triple product obtained by polarizing~(\ref{Det}).

\section{\textbf{The Jordan Formulation of Quantum Mechanics}}
\label{JQM}

In the Dirac formulation of quantum mechanics, a quantum mechanical state is
represented by a {\it complex\/} vector $v$, often written as $|v\rangle$,
which is usually normalized such that $v^\dagger v=1$.  In the Jordan
formulation~\cite{Jordan,JNW,GT}, the same state is instead represented by the
Hermitian matrix $vv^\dagger$, also written as $|v\rangle\langle{v}|$, which
squares to itself and has trace 1 (compare~(\ref{Cayley})).  The matrix
$vv^\dagger$ is thus the projection operator for the state $v$, which can also
be viewed as a pure state in the density matrix formulation of quantum
mechanics.  Note that the phase freedom in $v$ is no longer present in
$vv^\dagger$, which is uniquely determined by the state (and the normalization
condition).

A fundamental object in the Dirac formalism is the probability amplitude
$v^\dagger w$, or $\langle{v}|w\rangle$, which is not however measurable; it
is the the squared norm
$\left|\langle{v}|w\rangle\right|^2 = \langle{v}|w\rangle \langle{w}|v\rangle$
of the probability amplitude which yields the measurable transition
probabilities.  One of the basic observations which leads to the Jordan
formalism is that these transition probabilities can be expressed entirely in
terms of the Jordan product of projection operators, since
\begin{equation}
\label{TraceID}
(v^\dagger w)(w^\dagger v) \equiv \tr(vv^\dagger \circ ww^\dagger)
\end{equation}
A similar but less obvious translation scheme also exists~\cite{GT} for
transition probabilities of the form $|\langle{v}|A|w\rangle|^2$, where $A$ is
a Hermitian matrix, corresponding (in both formalisms) to an observable, so
that all {\it measurable\/} quantities in the Dirac formalism can be expressed
in the Jordan formalism.

So far, we have assumed that the state vector $v$ and the observable $A$ are
complex.  But the Jordan formulation of quantum mechanics uses only the {\it
Jordan identity}
\begin{equation}
\label{JID}
(A\circ B)\circ A^2 = A\circ \left(B\circ A^2\right)
\end{equation}
for 2 observables (Hermitian matrices) $A$ and $B$.  As shown in~\cite{JNW},
the Jordan identity~(\ref{JID}) is equivalent to power associativity, which
ensures that arbitrary powers of Jordan matrices --- and hence of quantum
mechanical observables --- are well-defined.

The Jordan identity~(\ref{JID}) is the defining property of a {\it Jordan
algebra}~\cite{Jordan}, and is clearly satisfied if the operator algebra is
associative, which will be the case if the elements of the Hermitian matrices
$A$, $B$ themselves lie in an associative algebra.  Remarkably, the only
further possibility is the Albert algebra of $3\times3$ octonionic Hermitian
matrices~\cite{JNW,Albert}.
\footnote{The $2\times2$ octonionic Hermitian matrices also form a Jordan
algebra, but, even though the octonions are not associative, it is possible to
find an associative algebra which leads to the same Jordan
algebra~\cite{JNW,Jacobson}.}
In what follows we will restrict our attention to this exceptional case.

\section{\textbf{The Jordan Eigenvalue Problem}}
\label{JORDAN}

Consider finally the eigenmatrix problem~(\ref{JEigen}).  Note first of all
that since $\AA$ and $\VV$ are Jordan matrices, the left-hand-side is
Hermitian, which forces $\lambda$ to be real.

Suppose first that $\AA$ is diagonal.  Then the diagonal elements $p$, $m$,
$n$ are clearly eigenvalues, with obvious diagonal eigenmatrices.  But there
are also other ``eigenvalues'', namely the averages $(p+m)/2$, $(m+n)/2$,
$(n+p)/2$.  However, the corresponding eigenmatrices --- which are related to
Peirce decompositions~\cite{Jacobson,Schafer} --- have only zeros on the
diagonal.  Thus, by~(\ref{NormZero}), they can not satisfy~(\ref{Cayley}), and
hence can not be written in the form~(\ref{ISq}).  To exclude this case, we
therefore restrict $\VV$ in (\ref{JEigen}) to the Cayley plane~(\ref{Cayley}),
which ensures that the eigenmatrices $\VV$ are primitive idempotents; they
really do correspond to ``eigenvectors'' $v$.  Recall that this forces the
components of $\VV$ to lie in a quaternionic subalgebra of $\OO$ (which
depends on $\VV$) even though the components of $\AA$ may not.

Next consider the characteristic equation
\begin{equation}
\label{Char}
-\det(\AA-\lambda\,\II)
  = \lambda^3 - (\tr\,\AA)\,\lambda^2 + \sigma(\AA)\,\lambda - (\det \AA)\,\II
  = 0
\end{equation}
It is not at first obvious that all solutions $\lambda$ of (\ref{Char}) are
real.  To see that this is indeed the case, we note that $\AA$ can be
rewritten as a $24\times24$ real symmetric matrix, whose eigenvalues are of
course real.  However, as discussed in~\cite{Eigen}, these latter eigenvalues
do {\it not\/} satisfy the characteristic equation~(\ref{Char})!  Rather, they
satisfy a modified characteristic equation of the form
\begin{equation}
\label{Cubic}
\det(\AA-\lambda\,\II) + r = 0
\end{equation}
where $r$ is either of the roots of a quadratic equation which depends on
$\AA$.  As shown explicitly using {\slshape Mathematica} in Figure~5
of~\cite{Find}, not only are these roots real, but they have opposite signs
(or at least one is zero).  But, as can be seen immediately using elementary
graphing techniques, if the cubic equation~(\ref{Cubic}) has 3 real roots for
both a positive and a negative value of $r$, it also has 3 real roots for all
values of $r$ in between, including $r=0$.  This shows that (\ref{Char}) does
indeed have 3 real roots.

Alternatively, since $F_4$ preserves both the determinant and the trace (and
therefore also~$\sigma$)~\cite{Freudenthal,Harvey}, it leaves the
characteristic equation invariant.  Since $F_4$ can be used to diagonalize
$\AA$~\cite{Freudenthal,Harvey}, and since the resulting diagonal elements
clearly satisfy the characteristic equation, we have another, indirect, proof
that the characteristic equation has 3 real roots.  Furthermore, this shows
that these roots correspond precisely to the 3 real eigenvalues whose
eigenmatrices lie in the Cayley plane.  We therefore reserve the word
``eigenvalue'' for the 3 solutions of the characteristic
equation~(\ref{Char}), explicitly excluding their averages.  The above
argument shows that these correspond to solutions $\VV$ of (\ref{JEigen})
which lie in the Cayley plane; we will verify this explicitly below.

Restricting the eigenvalues in this way corresponds to the traditional
eigenvalue problem in the following sense.  If $\AA$, $v\ne0$ lie in a
quaternionic subalgebra of the octonions, then the Jordan eigenvalue
problem~(\ref{JEigen}) together with the restriction~(\ref{Cayley}) becomes
\begin{equation}
\label{CEigen}
\AA \, vv^\dagger + vv^\dagger \AA = 2 \lambda \,vv^\dagger
\end{equation}
Multiplying~(\ref{CEigen}) on the right by $v$ and simplifying the result
using the trace of~(\ref{CEigen}) leads immediately to $Av=\lambda v$ (with
$\lambda\in\RR$), that is, the Jordan eigenvalue equation implies the ordinary
eigenvalue equation in this context.  Since the converse is immediate, the
Jordan eigenvalue problem~(\ref{JEigen}) (with $\VV$ restricted to the Cayley
plane but $\AA$ octonionic) is seen to be a reasonable generalization of the
ordinary eigenvalue problem.

We now show how to construct eigenmatrices $\VV$ of~(\ref{JEigen}), restricted
to lie in the Cayley plane, and with real eigenvalues $\lambda$ satisfying the
characteristic equation~(\ref{Char}).  From the definition of the determinant,
we have for real $\lambda$ satisfying~(\ref{Char})
\begin{equation}
0 = \det(\AA-\lambda\,\II) =
  (\AA-\lambda\,\II) \circ \Big( (\AA-\lambda\,\II)*(\AA-\lambda\,\II) \Big)
\end{equation}
Thus, setting
\begin{equation}
\QQ_\lambda = (\AA-\lambda\,\II)*(\AA-\lambda\,\II)
\end{equation}
we have
\begin{equation}
\label{QEigen}
(\AA-\lambda\,\II) \circ \QQ_\lambda = 0
\end{equation}
so that $\QQ_\lambda$ is a solution of~(\ref{JEigen}).

Due to the identity~(\ref{Springer}), we have
\begin{equation}
\label{QSq}
\QQ_\lambda * \QQ_\lambda=0
\end{equation}
If $\QQ_\lambda\ne0$, we can renormalize $\QQ_\lambda$ by defining
\begin{equation}
\PP_\lambda = {\QQ_\lambda\over\tr(\QQ_\lambda)}
\end{equation}
Each resulting $\PP_\lambda$ is in the Cayley plane, and is hence a primitive
idempotent.  Due to~(\ref{QSq}), we can write
\begin{equation}
\PP_\lambda=v_\lambda^{\phantom\dagger} v_\lambda^\dagger
\end{equation}
and we call $v_\lambda$ the (generalized) eigenvector of $\AA$ with eigenvalue
$\lambda$.  Note that $v_\lambda$ does {\it not\/} in general satisfy either
(\ref{Left}) or (\ref{Right}).  Rather, we have
\begin{equation}
\AA \circ v_\lambda^{\phantom\dagger} v_\lambda^\dagger
  = \lambda \, v_\lambda^{\phantom\dagger} v_\lambda^\dagger
\end{equation}
as well as
\begin{equation}
v_\lambda^\dagger v_\lambda^{\phantom\dagger} = 1
\end{equation}

Writing out all the terms and using (\ref{TID}) and (\ref{ID}), one computes
directly that
\begin{equation}
\QQ_\lambda\circ(\AA\circ \QQ_\mu) = (\QQ_\lambda\circ \AA)\circ \QQ_\mu
\end{equation}
If $\lambda$, $\mu$ are solutions of the characteristic equation (\ref{Char}),
then using~(\ref{QEigen}) leads to 
\begin{equation}
\mu\,(\QQ_\lambda\circ \QQ\mu) = \lambda\,(\QQ_\lambda\circ \QQ\mu)
\end{equation}
If we now assume $\lambda\ne\mu$ and $\QQ_\lambda\ne0\ne \QQ_\mu$, this shows
that eigenmatrices corresponding to different eigenvalues are orthogonal in
the sense
\begin{equation}
\PP_\lambda\circ \PP_\mu = 0
\end{equation}
where we have normalized the eigenmatrices.

We now turn to the case $\QQ_\lambda=0$.  We have first that
\begin{equation}
\label{QID}
\tr(\QQ_\lambda)
  = \tr \Big( (\AA-\lambda\,\II)*(\AA-\lambda\,\II) \Big)
  = \sigma(\AA-\lambda\,\II)
\end{equation}
Denoting the 3 real solutions of the characteristic equation~(\ref{Char}) by
$\lambda$, $\mu$, $\nu$, so that
\begin{eqnarray}
\tr\,\AA &=& \lambda + \mu + \nu \\
\sigma(\AA) &=& \lambda (\mu + \nu) + \mu\nu
\end{eqnarray}
we then have
\begin{equation}
\label{SigmaID}
\sigma(\AA-\lambda\,\II)
  = \sigma(\AA) - 2\lambda \, \tr\,\AA + 3\lambda^2
  = (\lambda - \mu) (\lambda - \nu)
\end{equation}
But by~(\ref{QSq}) and~(\ref{NormZero}), $\QQ_\lambda=0$ if and only if
$\tr(\QQ_\lambda)=0$.  Using~(\ref{QID}) and~(\ref{SigmaID}), we therefore see
that $\QQ_\lambda=0$ if and only if $\lambda$ is a solution of~(\ref{Char}) of
multiplicity greater than 1.  We will return to this case below.

Putting this all together, if there are no repeated solutions of the
characteristic equation~(\ref{Char}), then the eigenmatrix problem leads to
the decomposition
\begin{equation}
\label{Decomp}
\AA = \sum_{i=1}^3 \lambda_i \PP_{\lambda_i}
\end{equation}
in terms of orthogonal primitive idempotents, which expresses each Jordan
matrix $\AA$ as a sum of squares of {\it quaternionic\/} columns.
\footnote{To see this, one easily verifies that $\tr(\BB) = 0 = \sigma(\BB)$,
where $\BB=\AA-\sum \lambda_i \PP_{\lambda_i}$.  But this implies that
$\tr(\BB^2)=0$, which forces $\BB=0$.}
We emphasize that the components of the eigenmatrices $\PP_{\lambda_i}$ need
not lie in the same quaternionic subalgebra, and that $\AA$ is octonionic.
Nonetheless, it is remarkable that $\AA$ admits a decomposition in terms of
matrices which are, individually, quaternionic.

We now return to the case $\QQ_\lambda=0$, corresponding to repeated
eigenvalues.  If $\lambda$ is a solution of the characteristic
equation~(\ref{Char}) of multiplicity 3, then $\tr\,\AA=3\lambda$ and
$\sigma(\AA)=3\lambda^2$.  As shown in~\cite{Eigen} in a different context, or
using an argument along the lines of Footnote~\arabic{footnote}, this forces
$\AA=\lambda\,\II$, which has a trivial decomposition into orthonormal
primitive idempotents.  We are left with the case of multiplicity 2,
corresponding to $\AA\ne\lambda\,\II$ and $\QQ_\lambda=0$.

Since $\QQ_\lambda=0$, $\AA-\lambda\,\II$ is (up to normalization) in the
Cayley plane, and we have
\begin{equation}
\label{IISq}
\AA - \lambda\,\II = \pm ww^\dagger
\end{equation}
with the components of $w$ in some quaternionic subalgebra of $\OO$.  While
$ww^\dagger$ is indeed an eigenmatrix of $\AA$, it has eigenvalue
$\mu=\tr(\AA)-2\lambda\ne\lambda$.  However, it is straightforward to
construct a vector $v$ orthogonal to $w$ in a suitable sense.  For instance,
if
\begin{equation}
w = \pmatrix{x\cr y\cr r\cr}
\end{equation}
with $r\in\RR$, then choosing
\begin{equation}
v = \pmatrix{|y|^2\cr -y\bar{x}\cr 0\cr}
\end{equation}
leads to
\begin{equation}
\label{VWperp}
vv^\dagger \circ ww^\dagger = 0
\end{equation}
and only minor modifications are required to adapt this example to the general
case.  But (\ref{IISq}) now implies that
\begin{equation}
\AA \circ vv^\dagger = \lambda  \, vv^\dagger
\end{equation}
so that we have constructed an eigenmatrix of $\AA$ with eigenvalue $\lambda$.

We can now perturb $\AA$ slightly by adding $\epsilon\,vv^\dagger$, thus
changing the eigenvalue of $vv^\dagger$ by~$\epsilon$.  The resulting matrix
will have 3 unequal eigenvalues, and hence admit a decomposition
(\ref{Decomp}) in terms of orthogonal primitive idempotents.  But these
idempotents will also be eigenmatrices of $\AA$, and hence yield an orthogonal
primitive idempotent decomposition \hbox{of $\AA$}.~%
\footnote{More formally, with the above assumptions we have
\begin{equation}
\label{Exp}
(\AA+\epsilon\,vv^\dagger-\lambda\,\II)*(\AA+\epsilon\,vv^\dagger-\lambda\,\II)
  = (ww^\dagger+\epsilon\,vv^\dagger)*(ww^\dagger+\epsilon\,vv^\dagger)
  = 2 \epsilon\,vv^\dagger*ww^\dagger
\end{equation}
The Freudenthal square of (\ref{Exp}) is zero by (\ref{OCross}), which shows
that $\det(\AA+\epsilon\,vv^\dagger-\lambda\,\II)=0$ by~(\ref{Springer}), so
that $\lambda$ is indeed an eigenvalue of the perturbed matrix
$\AA+\epsilon\,vv^\dagger$.  Furthermore, (\ref{Exp}) itself is not zero
(unless $v$ or $w$ vanishes) since (\ref{VWperp}) implies that
\begin{equation}
2\, \tr(vv^\dagger*ww^\dagger) = (v^\dagger v) (w^\dagger w) \ne 0
\end{equation}
which shows that $\lambda$ does not have multiplicity 2.}
In summary, decompositions analogous to~(\ref{Decomp}) can also be found when
there is a repeated eigenvalue, but the terms corresponding to the repeated
eigenvalue can not be written in terms of the projections $\PP_\lambda$, and
of course the decomposition of the corresponding eigenspace is not unique.
\footnote{An invariant orthogonal idempotent decomposition when $\lambda$ is
an eigenvalue of multiplicity 2 is
\begin{equation}
\AA = \mu \, {(\AA-\lambda\,\II)\over\tr(\AA-\lambda\,\II)}
	- \lambda \, {\tilde{(\AA-\lambda\,\II)}\over\tr(\AA-\lambda\,\II)}
\end{equation}
where the coefficient of $\mu=\tr(\AA)-2\lambda$ is the primitive idempotent
corresponding to the other eigenvalue and the coefficient of $\lambda$ is an
idempotent but not primitive.  An equivalent expression was given
in~\cite{JNW}.}

\section{\textbf{Discussion}}
\label{DISCUSSION}

We have argued elsewhere~\cite{Dim,Spin} that the ordinary momentum-space
(massless and massive) Dirac equation in $3+1$ dimensions can be obtained via
dimensional reduction from the Weyl (massless Dirac) equation in $9+1$
dimensions.  This latter equation can be written as the eigenvalue problem
\begin{equation}
\label{Dirac}
\tilde{P}\psi = 0
\end{equation}
where $P$ is a $2\times2$ octonionic Hermitian matrix corresponding to the
10-dimensional momentum and tilde again denotes trace reversal.  The general
solution of this equation is
\begin{eqnarray}
P &=& \pm\theta\theta^\dagger \\
\psi &=& \theta\xi
\end{eqnarray}
where $\theta$ is a 2-component octonionic vector whose components lie in the
same complex subalgebra of $\OO$ as do those of $P$, and where $\xi\in\OO$ is
arbitrary.  (Such a $\theta$ must exist since $\det(P)=0$.)

It is then natural to introduce a 3-component formalism; this approach was
used by Schray~\cite{Schray,Thesis} for the superparticle.  Defining
\begin{equation}
\Psi = \pmatrix{\theta\cr \bar\xi}
\end{equation}
we have first of all that
\begin{equation}
\PP := \Psi \Psi^\dagger = \pmatrix{P& \psi\cr \psi^\dagger& |\xi|^2\cr}
\end{equation}
so that $\Psi$ combines the bosonic and fermionic degrees of freedom.  Lorentz
transformations can be constructed by iterating (``nesting'') transformations
of the form~\cite{Lorentz}
\begin{eqnarray}
P &\mapsto& MPM^\dagger \label{MTrans}\\
\psi &\mapsto& M\psi \label{VTrans}
\end{eqnarray}
which can be elegantly combined into the transformation
\begin{equation}
\label{Trans}
\PP \mapsto \MM\PP\MM^\dagger
\end{equation}
with
\begin{equation}
\MM = \pmatrix{M& 0\cr 0& 1\cr}
\end{equation}
This in fact shows how to view $SO(9,1)$ as a subgroup of $E_6$; the rotation
subgroup $SO(9)$ lies in $F_4$.  It turns out that the Dirac
equation~(\ref{Dirac}) is equivalent to the equation
\begin{equation}
\PP*\PP=0
\end{equation}
which shows both that solutions of the Dirac equation correspond to the Cayley
plane and that the Dirac equation admits $E_6$ as a symmetry group.  Using the
particle interpretation from~\cite{Dim,Spin} then leads to the interpretation
of (part of) the Cayley plane as representing 3 generations of leptons.

The modern description of symmetries in nature is in terms of Lie algebras.
For instance, one describes angular momentum by taking an infinitesimal
rotation, regarding it as a self-adjoint operator, and studying the resulting
eigenvalue problem.  Thus, if $A$ is the (self-adjoint version of the)
infinitesimal rotation $M$, then the rotation~(\ref{VTrans}) leads to the
eigenvalue problem $A\psi=\lambda\psi$.  But the infinitesimal form
of~(\ref{MTrans}) is essentially $A\circ P$, although in the octonionic case,
it is not clear how best to make $A$ self-adjoint.  It thus seems natural to
study the ($3\times3$) Jordan eigenvalue problem associated
with~(\ref{Trans}).

Finally, we refer to decompositions of the form~(\ref{Decomp}) as $p$-square
decompositions, where $p$ is the number of nonzero eigenvalues, and hence the
number of nonzero primitive idempotents in the decomposition.  If
$\det(\AA)\ne0$, then $\AA$ is a 3-square.  If $\det(\AA)=0\ne\sigma(\AA)$,
then $\AA$ is a 2-square.  Finally, if $\det(\AA)=0=\sigma(\AA)$, then $\AA$
is a 1-square (unless also $\tr(\AA)=0$, in which case $\AA\equiv0$).  It is
intriguing that, since $E_6$ preserves both the determinant and the condition
$\sigma(\AA)=0$, $E_6$ therefore preserves the class of $p$-squares for each
$p$.  If, as argued above, 1-squares correspond to leptons, is it possible
that 2-squares are mesons and 3-squares are baryons?

\newpage
\section*
  {\bfseries\boldmath APPENDIX: Diagonalizing Jordan Matrices Using $F_4$}

We start with a Jordan matrix in the form~(\ref{Three}), and show how to
diagonalize it using nested $F_4$ transformations.  As discussed in
\cite{Harvey}, a set of generators for $F_4$ can be obtained by considering
its $SO(9)$ subgroups, which in turn can be generated by $2\times2$ tracefree,
Hermitian, octonionic matrices.

Just as for the traditional diagonalization procedure, it is first necessary
to solve the characteristic equation for the eigenvalues.  Let $\lambda$ be a
solution of (\ref{Char}), and let $vv^\dagger\ne0$ be a solution of
(\ref{JEigen}) with eigenvalue $\lambda$.
\footnote{It is straightforward to construct $v$ using the results of
Section~\ref{JORDAN}, especially since we can assume without loss of
generality that $\lambda$ is an eigenvalue of multiplicity 1.}
We assume further that the phase in $v$ is chosen such that
\begin{equation}
v = \pmatrix{x\cr y\cr r\cr}
\end{equation}
where $x,y\in\OO$ and $r\in\RR$.  Define
\begin{equation}
\MM_1 = \pmatrix{-r& 0& x\cr
		0& N_1& 0\cr
		\bar{x}& 0& r\cr}
	  \Bigg/ N_1
\qquad
\MM_2 = \pmatrix{N_2& 0& 0\cr
		0& -N_1& y\cr
		0& \bar{y}& N_1\cr}
	  \Bigg/ N_2
\end{equation}
where the normalization constants are given by $N_1^2=|x|^2+r^2$ and
$N_2^2=N_1^2+|y|^2\equiv v^\dagger v\ne0$.  (If $N_1=0$, then $\AA$ is already
block diagonal.)  It is straightforward to check that
\begin{equation}
\MM_2 \MM_1 v = \pmatrix{0\cr 0\cr 1\cr}
\end{equation}
and, since everything so far is quaternionic, that
\begin{equation}
\MM_2 \MM_1 vv^\dagger \MM_1 \MM_2 = \pmatrix{0&0&0\cr 0&0&0\cr 0&0&1\cr}
  =: \EE_3
\end{equation}

But conjugation by each of the $\MM_i$ is an $F_4$ transformation (which is
well-defined since each $\MM_i$ separately has components which lie in a {\it
complex\/} subalgebra of $\OO$); this is precisely the form of the generators
referred to earlier.  Furthermore, $F_4$ is the automorphism group of the
Jordan product~(\ref{JProd}).  Thus, since
\begin{equation}
(\AA-\lambda\,vv^\dagger) \circ vv^\dagger = 0
\end{equation}
then after applying the (nested!)\ $F_4$ transformation above, we obtain
\begin{equation}
\Bigg( \MM_2 \Big(\MM_1 (\AA-\lambda\,\II) \MM_1\Big) \MM_2 \Bigg) \circ \EE_3
  = 0
\end{equation}
which in turn forces
\begin{equation}
\MM_2 (\MM_1 \AA \MM_1) \MM_2 = \pmatrix{X& 0\cr 0& \lambda\cr}
\end{equation}
where
\begin{equation}
X = \pmatrix{s& z\cr \bar{z}& t\cr}
\end{equation}
is a $2\times2$ octonionic Hermitian matrix (with $z\in\OO$ and $s,t\in\RR$).

The final step amounts to the diagonalization of $X$, which is easy.  Let
$\mu$ be any eigenvalue of $X$ (which in fact means that it is another
solution of (\ref{Char})) and set
\begin{equation}
\MM_3 = \pmatrix{\mu-t& 0& 0\cr
		0& t-\mu& z\cr
		0& \bar{z}& N_3\cr}
	  \Bigg/ N_3
\end{equation}
where $N_3=(\mu-t)^2+|z|^2$.  (If $N_3=0$, $X$ is already diagonal.)  This
finally results in
\begin{equation}
\MM_3 \Big(\MM_2 \left(\MM_1 \AA \MM_1\right) \MM_2\Big) \MM_3
  = \pmatrix{\mu& 0& 0\cr 0& \tr(X)-\mu& 0\cr 0& 0& \lambda\cr}
\end{equation}
and we have succeeded in diagonalizing $\AA$ using $F_4$ as claimed.

\section*{Acknowledgments}

CAM would particularly like to thank David Fairlie for having suggested to her
the relevance of the Jordan matrices many years ago.

\end{document}